\documentclass[11pt,a4paper]{article}
\usepackage{jheppub}

\usepackage{amsmath}
\usepackage{latexsym}

\title{\Large{Analytical and numerical analysis of a rotational invariant D=2 harmonic oscillator in the light of different noncommutative phase-space configurations}}
\author[a,b]{Everton M. C. Abreu,}
\author[b]{Mateus V. Marcial,}
\author[b]{Albert C. R. Mendes}
\author[b]{and Wilson Oliveira}

\affiliation[a]{Grupo de F\'isica Te\'orica, Departamento de F\'{\i}sica, \\
Universidade Federal Rural do Rio de Janeiro\\
BR 465 km 07, 23890-971, Serop\'edica, RJ, Brazil}
\affiliation[b]{Departamento de F\'{\i}sica, ICE, Universidade Federal de Juiz de Fora,\\
36036-330, Juiz de Fora, MG, Brazil}

\emailAdd{evertonabreu@ufrrj.br}
\emailAdd{mateusmarcial@fisica.ufjf.br}
\emailAdd{albert@fisica.ufjf.br}
\emailAdd{wilson@fisica.ufjf.br}
\abstract{In this work we have investigated some  properties of classical  phase-space  with symplectic structures consistent, at the classical level, with two noncommutative (NC)  algebras: the Doplicher-Fredenhagen-Roberts algebraic  relations and the NC approach which uses an extended Hilbert space  with rotational symmetry. This extended Hilbert space includes the  operators $\theta^{ij}$ and their conjugate momentum $\pi_{ij}$ operators.  In this scenario, the equations of motion for all extended phase-space coordinates with  their corresponding solutions were determined and a rotational invariant  NC  Newton's second law  was written.  As an application, we treated a NC harmonic oscillator constructed in this extended Hilbert space.   We have showed precisely that its solution is still periodic if and only if the ratio between the frequencies of oscillation  is a rational number. We investigated, analytically and numerically, the  solutions of this NC oscillator in a two-dimensional phase-space. The result led us to conclude that noncommutativity induces a stable perturbation into the  commutative standard oscillator and that the rotational symmetry is not broken.  Besides, we have demonstrated through the equations of motion that a zero momentum $\pi_{ij}$ originated a constant NC parameter, namely, $\theta^{ij}=const.$, which changes the original variable characteristic of $\theta^{ij}$ and reduces the phase-space of the system. This result shows that the momentum $\pi_{ij}$ is relevant and cannot be neglected when we have that $\theta^{ij}$ is a coordinate of the system. }


\usepackage{float}

\usepackage{graphicx}% Include figure files
\usepackage{dcolumn}% Align table columns on decimal point
\usepackage{bm}% bold math
\usepackage{amsmath}
\usepackage{latexsym}
\newcommand{\be}{\begin{equation}}
\newcommand{\ee}{\end{equation}}
\newcommand{\ea}{\end{eqnarray}}
\newcommand{\ba}{\begin{eqnarray*}}

\def\[{\left\lbrack}
\def\]{\right\rbrack}

\def\({\left(}
\def\){\right)}

\begin{document}

\maketitle
\flushbottom

\pagestyle{myheadings}
%\markright{Noncommutative cosmological models}
\markright{Analytical and numerical analysis of a rotational...}
% % % % % % % % % % % % % % % % % % % % % % % % % % % % % % % % % % % % % % % % % % %
% % % % % % % % % % % % % % % % % % % % % % % % % % % % % % % % % % % % % % % % % % % %
% % % % % % % % % % % % % % % % % % % % % %

%\newpage

%\setlength{\baselineskip} {20 pt}

\section{Introduction}

In the last years a remarkable   interest has been dedicated to the investigation of  the properties  of  noncommutative (NC) spaces. The motivation for these studies comes from different areas such  as renormalization \cite{1}, quantum gravity  \cite{2}, string theory \cite{3} and  quantum Hall effect \cite{4}. Regardless of the motivation, the fundamental consideration  is that the position operators satisfy the   non-trivial commutation relation
\begin{equation}\label{eq:EFNC}
[\hat{x}^\mu,\hat{x}^\nu]=i\hbar {\theta}^{\mu\nu}\,\,,
\end{equation}
where the antisymmetric matrix  $ {\theta}^{\mu \nu} $, which appears on the right-hand side of Eq.(\ref{eq:EFNC}), in general may depend on the coordinates and can be  understood through several paths. For example,   from  studies of open string  dynamics on a brane undergoing a constant antisymmetric background magnetic field, one finds a NC spacetime with  this kind of  structure,  where the matrix  on the right-hand side of Eq.(\ref{eq:EFNC}) is  constant. However, the constant  parameter $ {\theta}^{\mu \nu} $ provides a determined direction into the spacetime which does not affect the translational invariance whereas the Lorentz symmetry is 
broken \cite{5}. The  breaking of Lorentz symmetry causes serious problems like  the vacuum birefringence effect \cite{6} that was not  observed experimentally yet. 
 However, to choose the matrix $ \theta^{\mu\nu} $ as equal to a constant  has  also favorable consequences. In this case,  NC field theories (NCFT)  can be treated as deformations of the usual field theories, constructed by replacing in  the action of the models  the ordinary multiplication of fields by the Moyal-Weyl one \cite{7,8},  defined by
 \begin{equation}
 \label{primeira}
 \phi_1 (x)\star \phi_2 (x) \,=\,exp\left( {i\over 2} \theta^{\mu \nu}
 \partial^{x}_{\mu}\partial^{y}_{\nu} \right) \phi_1 (x) \phi_2 (y)\mid_{x=y}.
 \end{equation}
From the definition (\ref{primeira}), we can observe  that theories constructed through the Moyal-Weyl product are highly nonlocal.  This condition leads to the specific mixture of scales called in the current literature as the Ultraviolet/Infrared (UV/IR) mixing. Besides, these NC field theories (NCFT) show other difficulties such as nonrenormalizability and nonunitarity \cite{5}.
   
The right hand-side of Eq.(\ref{eq:EFNC}) can also be seen as a tensorial operator that commutes with the position operators. This type of spacetime was constructed from classical general relativity and quantum mechanics concepts by Doplicher, Fredenhagen and Roberts (DFR). It was showed by them that rotational but not Lorentz invariant theory can be recovered \cite{2,9}. In \cite{10}, it was shown how to construct  NC quantum theories that are dynamically invariant under rotations. This was  carried out  using the DFR algebra and promoting the object of noncommutativity (NCY) to an operator on an extended Hilbert space. Its canonical  conjugate momentum was also included since the NC parameter is now considered as a coordinate of this NC spacetime.  For a review the interested reader can see \cite{101}.
 
Extended NC spaces have also been  studied in a classical way \cite{11} and in cosmological  models coupled  to different types of perfect fluids  in  a   NC phase-space \cite{20}. In  the last one,  the authors obtained the Newton's second law  (NSL) assuming that the phase-space has a symplectic structure consistent with the commutation rules of NC quantum mechanics (NCQM). However this  NSL  in an NC space  breaks the  rotational symmetry

The paper is organized  as follows: In section \ref{section2},  we show how to obtain NSL  in a  rotational invariant NC space.  To accomplish this,  we  assume that the NC classical phase-space has  a symplectic structure consistent with the commutation rules provided in \cite{10} and we  write the equations of motion of all phase-space coordinates  including NC coordinate $\theta^{\mu\nu}$ and its canonical conjugate momentum  $ \pi^{\mu\nu}$.   In section \ref{section3}, we have constructed NSL in this extended Hilbert space for a generalized potential.
In section \ref{section4}, we  apply the new NC NSL to treat the harmonic oscillator (HO). In this case,  we exhibited the equations of motion for all phase-space coordinates with  their corresponding solutions. We have also analyzed   the periodicity conditions of these solutions.  In section 5,
specifically  for  the two-dimensional oscillator  we obtained the solutions depending just on the initial conditions. 
Numerically speaking,  we have shown these solutions graphically for the  rotational invariant NC HO and the usual, or commutative,  HO   under the same initial conditions. Consequently,  the NCY effects in the  trajectories of the two-dimensional NC HO    can  be directly visualized. Section  \ref{section6} is reserved for conclusions.

\section{Rotational Invariant Newton's Second Law }
\label{section2}

NC classical mechanics (NCCM) can be developed in a phase-space   which has a symplectic structure consistent with the commutation rules  of the NCQM in the  extended Hilbert space \cite{7}. Initially, we suppose the existence of a set of  symplectic variables: $\xi^{a}$ and $\Omega^{ab}$, with  $a,b,d=1,2, ...2n$. Thus, given  $F$ and $G$ functions of  symplectic variables $\xi^{a}$ and $ \Omega^{bd}$, we can  define  a generalized symplectic structure as \cite{12}, 
\begin{eqnarray}\label{eq:GG}
  {\{}F,G{\}}={\{}\xi^{a},\xi^{b}{\}}\frac{\partial F}{\partial\xi^{a}}\frac{\partial G}{\partial \xi^{b}}
&+& {\{}\xi^{a},\Omega^{bd}{\}}\frac{\partial F}{\partial\xi^{a}}\frac{\partial G}{\partial \Omega^{bd}} \nonumber \\
&+& {\{}\Omega^{bd},\xi^a {\}}\frac{\partial F}{\partial\Omega^{bd}}\frac{\partial G}{\partial \xi^a}
+ {\{}\Omega^{ac},\Omega^{bd}{\}}\frac{\partial F}{\partial\Omega^{ac}}\frac{\partial G}{\partial \Omega^{bd}}\,\,,
    \end{eqnarray}
where the Einstein's summation convention is understood.

     Given a  general   Hamiltonian $ H=H(\xi^{a}, \Omega^{ab})$, and using the symplectic structure defined  in (\ref{eq:GG}),  we can write  the equations of motion  as follows
     \begin{eqnarray}\label{eq:EMs}
       \dot{\xi}^{a}={\{}\xi^{a},H{\}},
       \end{eqnarray}
              \begin{eqnarray*}
       \dot{\Omega}^{ab}={\{}\Omega^{ab},H{\}}.
       \end{eqnarray*}
      Similarly, for any function $ F$ defined in this space we can write
   \begin{equation}\label{eq:EMsa1}
   \dot{F}\,=\,{\{}F,H{\}},
   \end{equation}
% \begin{eqnarray*}
%   \dot{G}^{ab}={\{}G^{ab},H{\}}.
%   \end{eqnarray*}
   Hence, we have constructed  a general symplectic structure which can be used  to obtain the equations of motion  for all phase-space variables.
 
The symplectic structure consistent with the commutation relations derived from NCQM given in Eqs. (\ref{eq:CC})-(\ref{eq:xp}) can be written directly as 
      \begin{equation}\label{eq:ES}
      {\{}x^{i},x^{j}{\}}=\theta^{ij}, \ \ \ \  {\{}x^{i},p_{j}{\}}=\delta^{i}_{j}, \ \ \ \  {\{}x^{i},\theta^{jk}{\}}=0, \ \ \ \ {\{}x^{i},\pi_{jk}{\}}=-\frac{1}{2}{\delta^{il}}_{jk} p_{l},
      \end{equation}
      \begin{eqnarray*}
           {\{}p_{i},p_{j}{\}}=0 ,\ \ \ \ \ {\{}p_{i},\theta^{jk}{\}}=0, \ \ \ \ \  {\{}p_{i},\pi_{jk}{\}}=0 ,
            \end{eqnarray*}
       \begin{eqnarray*}
       {\{}\theta^{ik},\theta^{jl}{\}}=0, \ \ \ \ \  {\{}\theta^{ik},\pi_{jl}{\}}={\delta^{ik}}_{jl}, \ \ \ \ \ {\{}\pi_{ik},\pi_{jl}{\}}=0.
       \end{eqnarray*}
      Here   $ \theta^{ij}$ and $\pi_{ij}$ are antisymmetric matrices.
       From the symplectic structure (\ref{eq:ES}), we can rewrite the Poisson bracket of two functions $F$ and  $G$ of the  symplectic variables $\xi^{a}$, $ \theta^{ij}$, and $\pi_{ij}$ given in Eq.(\ref{eq:GG}) as follows
              
      \begin{eqnarray}\label{eq:GGES}
        {\{}F,G{\}}&=&\theta^{ij}\frac{\partial F}{\partial x^{i}}\frac{\partial G}{\partial x^{j}}+ \bigg{(}\frac{\partial F}{\partial x^{i}}\frac{\partial G}{\partial p_{i}}-\frac{\partial F}{\partial p_{i}}\frac{\partial G}{\partial x^{i}}\bigg{)} \nonumber \\
&+&        \bigg{(}\frac{\partial F}{\partial \theta^{ij}}\frac{\partial G}{\partial \pi_{ij}}-\frac{\partial F}{\partial \pi_{ij}}\frac{\partial G}{\partial \theta^{ij}}\bigg{)} 
%-        \bigg{(}\frac{\partial F}{\partial \theta^{ij}}\frac{\partial G}{\partial \pi_{ji}}-\frac{\partial F}{\partial \pi_{ji}}\frac{\partial G}{\partial \theta^{ij}}\bigg{)}+\\\nonumber
+        \bigg{(}\frac{\partial F}{\partial x^{i}}\frac{\partial G}{\partial \pi_{ji}}
-\frac{\partial F}{\partial {\pi}^{ji}}\frac{\partial G}{\partial x_{i}}\bigg{)}\frac{p_{j}}{2}. 
%+        \bigg{(}\frac{\partial F}{\partial \pi_{ij}}\frac{\partial G}{\partial x^{i}}-\frac{\partial F}{\partial \pi_{ji}}\frac{\partial G}{\partial x^{i}}\bigg{)}\frac{p_{j}}{2}.\\\nonumber
        \end{eqnarray}

\noindent where the terms that have a zero bracket in the algebra (2.9) were omitted.  Notice that in Eq. (2.10) we are working with the extended DFR NCY.  To recover DFR algebra we have to make $\pi_{ij} =0$.
  
   At this point, we can construct, analogously to the general case above, the symplectic structure of interest, which must be consistent with  the commutation rules \cite{7} described above
   \begin{equation}\label{eq:CC}
    [\textbf{x}^{i},\textbf{p}_j]=i\delta^{i}_{j},\ \ \ \ \ [\theta^{ij},\pi_{kl}]=i(\delta^{i}_{k}\delta^{j}_{l}-\delta^{i}_{l}\delta^{j}_{k})=i{\delta^{ij}}_{kl},
    \end{equation}
              \begin{eqnarray}
    [\textbf{\textbf{p}}_{i},\pi_{ij}]=0, \ \ \ \ \ \ \   [\textbf{p}_{k},\pi_{ij}]=0,
    \end{eqnarray}
        \begin{eqnarray}\label{aaa000}
   [\textbf{x}^{i},\textbf{x}^{j}]=i\theta^{ij},\ \ \ \ \ \ \  [\textbf{x}^k,\theta^{ij}]=0,
   \end{eqnarray}
    \begin{eqnarray}\label{eq:xp}
             [\textbf{x}^{i},\pi_{kl}]=\frac{i}{2}{\delta^{ij}}_{kl}\textbf{p}_j.
              \end{eqnarray}
where $\textbf{x}^i$ is  the position operator and  $\theta^{ij}$ is the NC coordinate operator, while  $\textbf{p}^i$  and $\pi_{ij}$ are their canonical conjugate momenta operators, respectively. 
     It is important to stress,  that the last commutation relations were obtained as a solution of the  resulting equation from  the Jacobi identity   formed by the three  functions  $\textbf{x}^{i}$, $\textbf{x}^{j}$ and $\pi_{kl}$ given by
            \begin{eqnarray}
            [[\textbf{x}^{i},\pi_{kl}],\textbf{x}^{j}]+[[\textbf{x}^{j},\textbf{x}^{i}],\pi_{kl}]+[[\pi_{kl},\textbf{x}^{j}],\textbf{x}^{i}]=0, 
               \end{eqnarray}
               
               \begin{eqnarray*}
               [[\textbf{x}^{i},\pi_{kl}],\textbf{x}^{j}]-[[\textbf{x}^{j},\pi_{kl}],\textbf{x}^{i}]=-i{\delta^{ij}}_{kl}.
               \end{eqnarray*}
which was obtained using the first equation in (\ref{aaa000}) and the second one in (\ref{eq:CC}).  These operatorial relations closes the so-called extended DFR quantum algebra and the construction of the extended Hilbert space \cite{10,101}.

\section{An example: the generalized extended Hamiltonian}
\label{section3}

In order to apply the formalism developed in the last section let us analyze a rotational invariant Hamiltonian which shows the proper commutative limit \cite{171,181,26,27}.  Let us add an analogous kinetic term to a standard Hamiltonian of the form \cite{10}
      \begin{equation}\label{eq:HIR}
      H=\frac{\pi^{2}}{2\Lambda}+\frac{p^{2}}{2m}+V(x^{i},p_{j},\theta^{ij},\pi_{ij})\,\,.
       \end{equation}
We can see clearly in (\ref{eq:HIR}) that the generalized potential is a function of the extended NC phase-space.  The parameter $\Lambda$ has dimension of $(length)^{-3}$.
The equations of motion corresponding to the algebra in Eqs. (\ref{eq:ES}) can be determined directly from Eq. (2.10)
             \begin{eqnarray}\label{eq:A1}
      \dot{x}^{i}
      =\theta^{ij}\frac{\partial V}{\partial x^{j}}+\bigg{(}\frac{\partial V}{\partial p^{i}}+\frac{p^{i}}{m}\bigg{)}+\bigg{(}\frac{ \pi^{ji}}{\Lambda}+\frac{\partial{V}}{\partial{\pi_{ji}}}\bigg{)}p_{j},\\\nonumber
      \end{eqnarray}
      \begin{equation}\label{eq:A2}
      \dot{p}_{i}=-\frac{\partial{V}}{\partial x^{i}},
      \end{equation}
      \begin{equation}\label{eq:A3}
      \dot{\theta}^{ij}
      =2\frac{\pi^{ij}}{\Lambda}+2\frac{\partial V}{\partial\pi_{ij}},
      \end{equation}
      \begin{equation}\label{eq:A4}
      \dot{\pi}_{ij}
      =-2\frac{\partial V}{\partial\theta^{ij}}+\frac{\partial V}{\partial x^{i}}p_{j}\,\,,
      \end{equation}
we can see clearly that, if $\pi_{ij}=0$ and consequently the potential would not be a function of $\pi_{ij}$, from Eq. (\ref{eq:A3}) that $\theta = const.$ and we recover the canonical NCY.  We will talk more about this result in the near future.

Substituting Eqs.(\ref{eq:A2})-(\ref{eq:A4}) into  the   derivative of Eq.(\ref{eq:A1})  with respect to time, namely $\ddot{x}\,=\,\{\dot{x}, H \}$, we  obtain that
                  \begin{eqnarray}\label{eq:2LNM}
        m\ddot{x}^{i}  &=&-\frac{\partial{V}}{\partial x_{i}}+m\bigg{[} \theta^{ij} \frac{\partial}{\partial x^{k}} \bigg{(}\frac{\partial V}{\partial x^{j}}\bigg{)}+p_j\frac{\partial}{\partial x^{k}} \bigg{(}\frac{\partial V}{\partial \pi_{ji}}\bigg{)}  +\frac{\partial}{\partial x^{k}} \bigg{(}\frac{\partial V}{\partial p^{i}}\bigg{)}\bigg{]}\dot{x}^{k}\nonumber \\
&+&        m\bigg{[}\frac{\partial}{\partial p_{k}} \bigg{(}\frac{\partial V}{\partial x^{j}}\bigg{)}\theta^{ij} + \bigg{(}\frac{\pi^{ki}}{\Lambda}+\frac{\partial{V}}{\partial{\pi_{ki}}}\bigg{)} + \frac{\partial}{\partial p_{k}} \bigg{(}\frac{\partial V}{\partial \pi_{ji}}\bigg{)}p_{j}+ \frac{\partial}{\partial p^{k}} \bigg{(}\frac{\partial V}{\partial p^{i}}\bigg{)}\bigg{]}\dot{p}_{k} \nonumber \\
&+&          m\bigg{[} \delta^{i}_{k}\delta^{j}_{l}\frac{\partial V}{\partial x^{j}}+\frac{\partial}{\partial \theta^{kl}} \bigg{(}\frac{\partial V}{\partial x^{j}}\bigg{)}\theta^{ij}+ \frac{\partial}{\partial \theta^{kl}} \bigg{(}\frac{\partial V}{\partial \pi_{ji}}\bigg{)}p_{j}+\frac{\partial}{\partial \theta^{kl}} \bigg{(}\frac{\partial V}{\partial p^{i}}\bigg{)}\bigg{]}\dot{\theta}^{kl} \nonumber \\
&+&            m\bigg{[} \frac{\partial}{\partial \pi_{kl}} \bigg{(}\frac{\partial V}{\partial x^{j}}\bigg{)}+ \frac{p_{j}}{\Lambda}\delta^{j}_{k}\delta^{i}_{l}+\frac{\partial}{\partial \pi_{kl}} \bigg{(}\frac{\partial V}{\partial \pi_{ji}}\bigg{)} +\frac{\partial}{\partial \pi_{kl}} \bigg{(}\frac{\partial V}{\partial p^{i}}\bigg{)}\bigg{]}\dot{\pi}_{kl}.
         \end{eqnarray}
      This equation is  the rotational invariant NSL  in an  extended NC  phase-space. In this equation,  the  corrections  due to the  NCY formalism with  the symplectic  structure (\ref {eq:ES}) are represented by  the terms  at the right-hand side of Eq. (\ref{eq:2LNM}), except  the first one.  All these new terms are generated by   the variations in the potential and  by the presence  of NC coordinates and its canonical conjugate momenta, which were crucial  to extend the  NC  phase-space and  also to  recover its  invariance under rotations.
      
The first two terms of Eq.(\ref{eq:2LNM}) were obtained in \cite {11} where the authors described a  NSL which is non-invariant  under rotations  in an NC phase-space but $\theta^{ij}$ was considered simply a   constant NC parameter,  while in our case $ \theta^{ij} $ is considered as a dynamic variable. Therefore, this new  rotational invariant NC  NSL, Eq.(\ref{eq:2LNM}), generalizes the results obtained in \cite{11,13}. 

Notice that if we make $\pi_{ij}=0$ in Eq. (3.1), all the $\pi_{ij}$ derivatives in Eqs. (3.2)-(3.6) disappear and consequently we recover the Doplicher-Fredenhagen-Roberts NC structure.  If we make $\theta^{ij}=\pi_{ij}=0$ we recover the commutative standard framework, of course.
   
\section{  Isotropic D-dimensional Noncommutative  Harmonic Oscillator }
\label{section4}

We will now treat an isotropic D-dimensional NC  HO (NCHO) which can be described by   the rotational invariant NC  Hamiltonian in Eq. (\ref{eq:HIR}) that has an appropriate commutative limit \cite{10} and
%  \begin{equation}\label{eq:HIRa1}
%       H=\frac{\pi^{2}}{2\Lambda}+\frac{p^{2}}{2m}+V(x^{i},p_{j},\theta^{ij},\pi_{ij})\,\,,
%        \end{equation}
where  the potential $V(x^{i},p_{j},\theta^{ij},\pi_{ij})$ will be given by
 \begin{equation}\label{vnc}
 V(x^{i},p_{j},\theta^{ij},\pi_{ij})=\frac{1}{2}m\omega^{2}\bigg{(}x^{i}+\frac{1}{2}\theta^{ij}p_j\bigg{)}^{2}+\frac{1}{2}\Lambda\Omega^{2}\theta^{2}.
 \end{equation}
where $\theta^2=\theta^{ij}\,\theta_{ij}$ which is a scalar constructed from the rank 2 tensor $\theta^{ij}$.
 By considering  this potential  we can rewrite Eqs. (\ref{eq:A1})-(\ref{eq:A4}) as
 
 \begin{eqnarray}\label{eq:A11}
 \dot{x}^{i}
 =\frac{1}{2}\theta^{ij}\bigg{(}m\omega^{2}x_{j}+\frac{1}{2}m\omega^{2}\theta_{jl}p^{l}\bigg{)}+\frac{p^{i}}{m}+\bigg{(}\frac{ \pi^{ji}}{\Lambda}\bigg{)}p_{j},\\\nonumber
 \end{eqnarray}
 \begin{equation}\label{eq:A22}
 \dot{p}_{i}=-m\omega^{2}x_{i}-\frac{1}{2}m\omega^{2}\theta_{ij}p^j,
 \end{equation}
 \begin{equation}\label{eq:A33}
 \dot{\theta}^{ij}=2\frac{\pi^{ij}}{\Lambda},
 \end{equation}
 \begin{eqnarray}\label{eq:A44}
 \dot{\pi}_{ij}
 =-2\Lambda\Omega^{2}\theta_{ij}.
 \end{eqnarray}

%{eq:A33}) 
Let us analyze these equations of motion in the light of DFR approach.  In a naive way, it would be possible that we think that when $\pi_{ij}=0$ it would be natural to conclude that the resulting phase-space would be given by the DFR one.  However, as we mentioned before when we have analyzed the equations of motion for $\theta^{ij}$ and 
$\pi_{ij}$ in Eqs. (3.4) and (3.5), the result obtained is that $\theta^{ij}=const.$.  And again, if $\pi_{ij}=0$ in (\ref{eq:A33}) we can see clearly that $\theta^{ij}=const..$
If we construct a Hamiltonian independent of $\pi_{ij}$ it does not make sense to construct Eqs. (3.5) and (4.6).  Substituting these values in Eqs. (4.3) and (4.4) we recover the canonical commutativity and not the DFR NCY approach.

Consequently we can conclude that the extended DFR and pure DFR formalisms are both connected to the cannonical NCY via $\pi_{ij}$ and not only via the nature of $\theta^{ij}$.  Namely, to promote a dimensional reduction of the phase-space (doing $\pi_{ij}=0$) means that $\theta^{ij}$ loses its variable parameter characteristic and becomes again a constant parameter.  Hence, the reduction is represented by $(x^i,p_i,\theta^{ij},\pi_{ij})\:\longrightarrow\: (x^i,p_i)$ where $\theta^{ij}$ is only a constant parameter, the result of the bracket between $x$'s.  

So, concerning the original DFR formalism, although in general, the momentum $\pi_{ij}$ may not be relevant, we understand that the momentum associated to $\theta^{ij}$ is necessary.  As a matter of fact, it would be natural and direct to construct this object since $\theta^{ij}$, in DFR phase-space, is a coordinate and must have an associated momentum.  However, what is new, in our point of view, is to connect the existence of $\pi_{ij}$ with the kind of the NCY or, in other words, if the NCY is DFR-extended or canonical.

This result make us think that, if we consider, for example, quantum field theories systems embedded in a NC spacetime, the  implications are even more serious because the existence of a variable NC parameter $\theta^{\mu\nu}$ recovers the Lorentz invariance of the NC theory.  But, the relevance of $\pi_{\mu\nu}=0$ is the fact that it brings back a constant $\theta^{\mu\nu}$, and hence we have the Lorentz invariance violated.  So, having said that, the connection between both objects ($\theta^{\mu\nu}$ and $\pi_{\mu\nu}$) is a connection between Lorentz invariant or non-invariant NC theories.
%$\pi=0 \rightarrow \theta=const.$ which represents the DFR NC phase-space, as we commented in the last section.  Since $\pi$ does not exist, it does not make sense to obtain (3.6).  Therefore, there is no contradiction between the value $\pi=0$ and Eq. (3.6).  The vice-versa effect works, namely, if we have $\theta=const.$ in Eq. (3.5) we will have $\pi=0$ and consequently (3.6) does not make sense.
To consider a similar analysis concerning anisotropic oscillators \cite{horvathy} would be interesting.

\subsection{A $\theta^{ij}$ equation for the noncommutative harmonic oscillator}

An interesting result is the one obtained when we substitute  Eq.(\ref{eq:A44}) into Eq.(\ref{eq:A33}) and one obtains a differential equation for the NC coordinate $ \theta^{ij} $, which is analogous   to the commutative  HO differential equation
 \begin{equation}
 \ddot{\theta}^{ij}+4\Omega^{2}\theta^{ij}=0\,\,,
 \end{equation}
and this $\theta$-HO differential equation shows that Eq. (3.6) is the expected one in the NC phase-space since we are working with the NCHO in Eqs. (3.1)-(3.2).

 A general solution of this last equation can be written as \footnote {Repeated indices do not indicate summation in terms like $ D^{ij}\cos(2 \omega t + \phi^{ij}) $ and $ D^{ij} \sin(2\omega t + \phi^{ij}) $.}
 
 \begin{equation}\label{eq:ET}
 \theta^{ij}(t)=D^{ij}\cos(2\Omega t+\phi^{ij})\,\,,
 \end{equation}
 where  $D^{ij}$ and $\phi^{ij}$ are constants. Inserting the value of $\theta^{ij}(t)$ into Eq.(\ref{eq:A33}) we have that
 \begin{equation}\label{eq:Epi}
 \pi_{ij}(t)=-\Lambda\Omega D^{ij}\sin(2\Omega t+\phi^{ij}).
 \end{equation}
%Performing  a coordinates transformations in phase-space analogous to  the definition  of  the shifted coordinate operator,  which is used in NC  quantum mechanic (NCQM) \cite{14,15,16,17}, defined as 
%        \begin{equation}\label{eq:TN}
%             {X}^{i}={x}^{i}+\frac{1}{2}\theta^{ij}p_j.
%        \end{equation}
%                Equations (\ref{eq:A11}) and (\ref{eq:A22}) will be immensely  simplified, resulting in the following solutions
and substituting Eq. (4.8) and (4.9) in Eq. (4.3) we have the solution for $x^i(\tau)$, which is 
              \begin{equation}\label{eq:sf}
                       x^{i}(t)=A^{i}\cos(\omega t+\phi^{i})+\frac{1}{2}\bigg{(}D^{ij}\cos(2\Omega t+\phi^{ij})\bigg{)}\bigg{[}m\omega A_{j}\sin(\omega t+\phi_{j})\bigg{]},
                       \end{equation}
and substituting Eq. (4.8) in Eq. (4.4) we can solve the differential equation for $p_i$ and                
                 \begin{equation}
                       p_{i}(t)=-m\omega A_{i}\sin(\omega t+\phi^{i})\,\,,
                       \end{equation}
                where $A_{i}$ and $\phi^{i}$ are constants.

%Notice that, of course, since we are working in a $D=6 \,(3+3)$ configuration space, the fact that we do not have any NC coordinate inside Eq. (4.10) can show us that in fact, when we solved the system (4.3)-(4.6) we have performed a dimensional reduction in this configuration space and now, Eq. (4.10) represents a $D=3$ solution for $x=x(t)$.

We can see clearly that the first term in the r.h.s. of Eq. (4.10) is the standard solution for a HO expected form.  The others terms are consequences of the NC characteristic of the phase-space, as we have explained so far.

%   By performing a transformation in phase-space analogous to  the definition  of  the shifted coordinate operator,  which is used in NC  quantum mechanic (NCQM) \cite{14,15,16,17} definided as 
%       \begin{equation}\label{eq:TN}
%            {X}^{i}={x}^{i}+\frac{1}{2}\theta^{ij}p_j,
%       \end{equation}
%     the Eqs.(\ref{eq:A11}) and (\ref{eq:A22})  may be written as
%      \begin{eqnarray}\label{eq:A111}
%      {\dot{X}}^{i}
%      =\frac{p^{i}}{m},
%      \end{eqnarray}
%      \begin{equation}\label{eq:A222}
%      \dot{p}_{i}=-m\omega^{2}X_{i}.
%      \end{equation}
%       Substitution of  Eq.(\ref{eq:A222})  in the time derivative of  Eq.(\ref{eq:A111})  yields
%      \begin{eqnarray}\label{eq:epx}
%      {\ddot{X}}^{i}+\omega^{2}X^{i}=0.
%      \end{eqnarray}
%     This equation admits the following solution
%      \begin{equation}
%      X^{i}(t)=A^{i}cos(\omega t+\phi^{i}).
%      \end{equation}
%      Where $A^{i}$ and $\phi^{i}$ are constants. From Eq.(\ref{eq:A111}) we find immediately that
%       \begin{equation}
%       p^{i}(t)=-m\omega A^{i}sen(\omega t+\phi^{i}).
%       \end{equation}
%      Finally, using the transformation (\ref{eq:TN})  we can  write   the following  solution for the position coordinates 
%             \begin{equation}\label{eq:sf}
%       x^{i}(t)=A^{i}cos(\omega t+\phi^{i})+\frac{1}{2}\bigg{(}D^{ij}cos(2\Omega t+\phi^{ij})\bigg{)}\bigg{[}m\omega A_{j}sen(\omega t+\phi_{j})\bigg{]}.
%       \end{equation}

        The second term on the  right-hand side of Eq. (\ref{eq:sf}) is the  corrections of the position coordinates associated with the NC symplectic structure  of the phase-space. This correction depends basically on the  NC coordinate  $\theta^{ij} $ and  the linear momentum $p^{i}$ of the system.
               
\subsection{ Periodicity of  NC Harmonic Oscillator }\label{APP}   

            Let us discuss the  periodicity conditions  of  the NC HO solution  given by Eq. (\ref{eq:sf}). For this purpose, let us assume that the HO solution $x^{i}(t)$ is periodic and has the period $T$. This  assumption  is true if the solution $x^{i}(t)$  satisfies the equation
            \begin{equation}\label{cpoh}
                    x^{i}(t+T)=x^{i}(t),\forall t\in    \mathbb{R}.
                    \end{equation}
                       From Eq. (\ref{eq:sf}) we can rewrite Eq. (\ref{cpoh})  as
                          \begin{eqnarray}\label{cpoh1}
&&x^{i}(t+T)-x^{i}(t)=A^{i}\cos(\omega t+\phi^{i}+\omega T)+\frac{1}{2}m\omega A_{j}D^{ij}\cos(2\Omega t +\phi^{ij}+2\Omega T)\nonumber \\
        &\times& \sin(\omega t+\phi_j+\omega T)-A^{i}\cos(\omega t+\phi^{i})-\frac{1}{2}m\omega A_jD^{ij}\bigg{(}\cos(2\Omega t+\phi^{ij})\bigg{)}\nonumber \\
        &\times& \sin(\omega t+\phi_{j})
        =A^{i}\cos(\omega t+\phi^{i})\cos(\omega T)-A^{i}\sin(\omega t +\phi^{i})\sin(\omega T)\nonumber \\
        &+&\frac{1}{2}m\omega A_{j}D^{ij}
        \bigg{[}\cos(2\Omega t+\phi^{ij})\cos(2\Omega T)
        -\sin(2\Omega t+\phi^{ij})\sin(2\Omega T)\bigg{]} \\
        &\times&\bigg{[}\sin(\omega t+\phi_j)\cos(\omega T)
        +\cos(\omega t +\phi_j)\sin(\omega T)\bigg{]}-A^{i}\cos(\omega t+\phi^{i}) \nonumber \\
        &-&\frac{1}{2}m\omega A_jD^{ij}\bigg{(}\cos(2\Omega t+\phi^{ij})\bigg{)}
        \sin(\omega t+\phi_{j})=0. \nonumber
         \end{eqnarray}
          %Só para tese%        
                          
The conditions  given by Eq.(\ref{cpoh1}),   thereafter Eq.(\ref{cpoh}), are satisfied if $\cos(\omega T)= 1$ and $ \cos (2 \Omega T) =1$. Then we can write  $\omega T=2\pi k$ and  $2\Omega T=2\pi l$, where $k$ and $l$ are  positive integers. Thus, the following condition  $$ \frac{2\Omega}{\omega}=\frac{l}{k}$$  is satisfied by the  frequencies  $\omega$ and  $\Omega$.
                          
Therefore, if $ \omega $ and $ \Omega $ are such that  $ 2 \Omega / \omega $ is a rational number,  for $$\frac {p}  {q}=\frac {2\, \Omega} {\omega}\,\,,$$ where numbers $ p $ and $ q $  are {\it relatively prime}.
%, i.e., $gcd (p, q) = 1$; 
The solution given by Eq.(\ref{eq:sf}) is  periodic with the  period given by
                                        \begin{equation}\label{CP1}
                                          T =\frac {2 \pi} {\omega} m, 
                                         \end{equation}
where $ m $ is the smallest positive integer such that $ \frac {p}{q}m $  is an integer, hence we can conclude that $ m = q $.                                        
In fact, as $ gcd (p, q) = 1 $ then by the  Bezout's identity there are integers $a$ and $b$ such that $ap + bq = 1$ then 
$$ m=map+mbq= \Big[\Big(\frac{p}{q} m \Big)a\,+\,mb\Big]q\,\,,$$ 
namely, $m>0$  is a multiple of $q$ and hence $ m\geq q $, which justifies the above conclusion. From  equation $$T=\frac{2\pi}{\omega}q\,\,,$$ we observe that if  $2\,\Omega $ and $ \omega $ are integers {\it relatively prime}, then the  period of oscillation is equal to $ 2\pi$, while for $\omega=2\,\Omega$ the period of NC HO is given by $T=2\pi/ \omega$, which coincides with the period of the commutative  HO.
%            
%      
%   Insertin the expression into eeee we obtain
%   
%   Within the contex
%Let us discuss about  periodicity conditions

\section{ $2$-D Noncommutative Harmonic Oscillator}
\label{section5}

We now wish to investigate  the  dynamics   of  the  $2$-D NCHO, using  the results  obtained in the previous section. Thus, according to Eq.(\ref{eq:sf}),  the solution
for the position coordinates   of the two-dimensional NC HO  can be written as
\begin{eqnarray}\label{poor}
 x^{1}(t)&=&A^{1}\cos(\omega t+\phi^{1})+\frac{1}{2}m \omega  D^{12}A_{2}\cos{(2\Omega t)}\sin(\omega t+\phi_{2}),\\\nonumber
  x^{2}(t)&=&A^{2}\cos(\omega t+\phi^{2})-\frac{1}{2}m \omega  D^{12}A_{1}\cos{(2\Omega t)}\sin( \omega t+\phi_{1}),\\\nonumber
    \end{eqnarray}
   since  the phase angle $ \phi^{ij} $ was  considered zero.
   
   Now  our main objective  is to show the effects of the NC  corrections in  the dynamical   behavior of a two-dimensional NCHO in an extended phase-space. This may be accomplished through a detailed comparison  between  the   solutions and  the trajectories   of   the  NC and  commutative    oscillators.  However, for this comparison  to be appropriate we need to have the freedom to choose the same initial configurations for both oscillators. In other words, we need to obtain the dynamical solutions    depending only  on  the NC coordinate,  the  initial positions and the initial velocities. Therefore, we consider that the initial conditions of the two oscillators are the initial position $ x^{i}_{00} $ and the initial velocity $ \dot{x}^{i}_{00} $. Thus, for the usual HO, indicated by the subscript $ 0 $, the desired solutions are
        \begin{eqnarray}\label{ioia1}
            x^{1}_{0}(t) &=&x^{1}_{00}\cos(\omega t)+\frac{\dot{x}^{1}_{00}}{\omega}\sin(\omega t),\nonumber \\
               x^{2}_{0}(t) &=&x^{2}_{00}\cos(\omega t)+\frac{\dot{x}^{2}_{00}}{\omega}\sin(\omega t).
           \end{eqnarray}
Hence, for the case of the NC oscillator, using Eq.(\ref{poor}), the solutions are given by
      \begin{eqnarray}\label{eq:sf33}
     &&  x^{1}(t)
           =\bigg{\{}\frac{\frac{1}{2} m D^{12} \dot{x}^{2}_{00}+ x^{1}_{00}}{1-(\frac{1}{2}m\omega D^{12})^{2}}\bigg{\}}\cos(\omega t)
           -\frac{1}{\omega}\bigg{\{}\frac{-\frac{1}{2}m \omega^{2}  D^{12} x^{2}_{00}+ \dot{x}^{1}_{00}}{(\frac{1}{2}m\omega D^{12})^{2}-1}\bigg{\}}\sin(\omega t)\\
            &+&\frac{1}{2}m\omega D^{12}\bigg{\{}\bigg{(}\frac{-\frac{1}{2}m   D^{12} \dot{x}^{1}_{00}+ {x}^{2}_{00}}{1-(\frac{1}{2}m\omega D^{12})^{2}}\bigg{)} \sin(\omega t)+
           \frac{1}{\omega}\bigg{(}\frac{\frac{1}{2}m \omega^{2}  D^{12} {x}^{1}_{00}+ \dot{x}^{2}_{00}}{(\frac{1}{2}m\omega D^{12})^{2}-1}\bigg{)}\cos(\omega t)\bigg{\}}\cos(2\Omega t),\nonumber
                    \end{eqnarray}
\noindent and
              \begin{eqnarray}\label{eq:sf4}
       &&    x^{2}(t)
               =\bigg{\{}\frac{-\frac{1}{2}m   D^{12} \dot{x}^{1}_{00}+ {x}^{2}_{00}}{1-(\frac{1}{2}m\omega D^{12})^{2}}\bigg{\}}\cos(\omega t)
               -\frac{1}{\omega}\bigg{\{}\frac{\frac{1}{2}m \omega^{2}  D^{12} {x}^{1}_{00}+ \dot{x}^{2}_{00}}{(\frac{1}{2}m\omega D^{12})^{2}-1}\bigg{\}}\sin(\omega t)\\
                &-&\frac{1}{2}m\omega D^{12}\bigg{\{}\bigg{(}\frac{\frac{1}{2} m D^{12} \dot{x}^{2}_{00}+ x^{1}_{00}}{1-(\frac{1}{2}m\omega D^{12})^{2}}\bigg{)} \sin(\omega t)+
               \frac{1}{\omega}\bigg{(}\frac{-\frac{1}{2}m \omega^{2}  D^{12} x^{2}_{00}+ \dot{x}^{1}_{00}}{(\frac{1}{2}m\omega D^{12})^{2}-1}\bigg{)}\cos(\omega t)\bigg{\}}\cos(2\Omega t), \nonumber
               \end{eqnarray}
where   $|\frac{1}{2}m\omega D^{12}|\ne 1 $.                     
                    Notice that, if we consider $\Omega =0$ in these equations, we obtain the  solutions of the usual   oscillator as a particular case.                                         
       Thus, we have  determined the dynamical solutions of the oscillator coordinates that are independent of their  phase angles and amplitude of  oscillation.
       
       Now  let us    consider the following initial conditions    $ {x}^{2}_{00} = 0$, $ {\dot{x}}^{1}_{00} = 0 $ and $ {\dot {x}}^{2}_{00}=0$. Thus,  using  Eqs.(5.3) and (5.4), for the   NCHO we have
       
                             \begin{eqnarray}\label{eq:sf333}
                                                  x^{1}(t)&=&\frac{x^{1}_{00}}{1-(\alpha\omega)^{2}}
                                                  \bigg{\{}1- (\alpha\omega)^{2} \cos(2\Omega t)\bigg{\}}  \cos(\omega t),        
                                                               \end{eqnarray}
               \begin{eqnarray}\label{eq:sf444}
                                                                x^{2}(t)&=&\frac{\alpha\omega x^{1}_{00}}{1-(\alpha\omega)^{2}}
                                                                \bigg{\{}1-  \cos(2\Omega t)\bigg{\}}  \sin(\omega t),         
                                                                             \end{eqnarray}
%        %             Finally, the trajectories  of the rotational invariant  NC HO can be visulized  in the graphics  of the  solutions given by Eqs.(\ref{eq:sf33}) and (\ref{eq:sf4}). In addition, we can  plot the graph of the usual oscillator  solutions,  with the same initial configuration. That will  easily show  the influences of NCY in the trajectories of the oscillators. To accomplish this,  let us    consider the following initial conditions   $ {x}^{1}_{00} = 1 $, $ {x}^{2}_{00} = 0$, $ {\dot{x}}^{1}_{00} = 1 $ and $ {\dot {x}}^{2}_{00}=0$. Thus, for the   NC HO we get
%                            \begin{eqnarray}\label{eq:sf333}
%                             x^{1}(t)&=&\bigg{\{}\frac{ 1}{1-(\alpha\omega)^{2}}\bigg{\}}cos(\omega t)
%                                 + \bigg{\{}  \frac{\alpha^{2} \omega^{2}  }{(\alpha\omega )^{2}-1}\bigg{\}}cos(\omega t)cos(2\Omega t),\\\nonumber
%                                          \end{eqnarray}
%                                                          \begin{eqnarray}\label{eq:sf444}
%                                 x^{2}(t) &=&
%                                     -\bigg{\{}\frac{\alpha \omega  }{(\alpha\omega )^{2}-1}\bigg{\}}sen(\omega t)-
%                                      \bigg{\{} \frac{\alpha \omega}{1-(\alpha\omega )^{2}}\bigg{\}} sen(\omega t)cos(2\Omega t).\\\nonumber
%                                              \end{eqnarray}
where we  have used   $\alpha=(\frac{1}{2}m D^{12})$. While, from Eq. (5.2),  the commutative  HO solution is given by
                       
                                            \begin{eqnarray}\label{ioi}
                                  x^{1}_{0}(t) &=&x^{1}_{00}\cos(\omega t),\\
                                     x^{2}_{0}(t) &=& 0. \nonumber
                                 \end{eqnarray}
                                 
\begin{figure}[H]
                                 \begin{center}
                             \includegraphics[width=2.31in,height=2.22in]{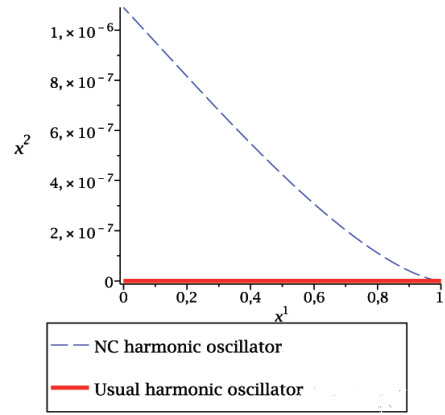}
                                   \end{center}
                                 \caption{For $t \leq \frac{\pi}{20}$}\label{fig1}
                               \end{figure}

                               Equations (\ref{eq:sf333}) and (\ref{eq:sf444}) tell us that the NC HO 
                              passes through the $x^{2}$-axis at the same instant of time as the usual HO passes through the origin. Furthermore, these are the only moments at which the NC HO crosses the $x^{2}$-axis provided that $|\alpha \omega|<1$. To exemplify this case, considering  $\alpha=10^{-6}$, $\omega =10$, and $\Omega= 3/2$ in Eqs.(\ref{eq:sf333}), (\ref{eq:sf444}), and  (\ref{ioi}), the trajectory of usual and NC HO is represented in figure (\ref{fig1}).

\begin{figure}[H]
                                       \begin{center}
                                       \includegraphics[width=2.31in,height=2.22in]{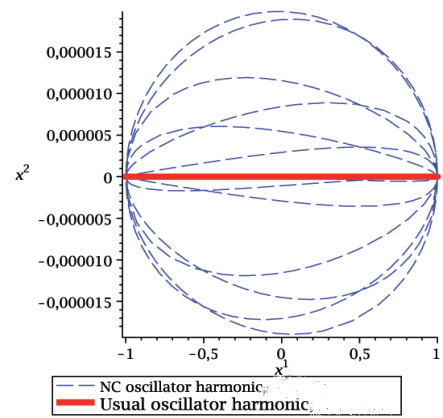}
                                            %\caption{ Position $x^{1}$ \textit{versus} $x^{2}$,  for  $t={2\pi}$.}
                                           \end{center}
                                        \caption{For $t \leq 2\,\pi$}\label{fig2}
                               \end{figure}

\begin{figure}[H]
                                               \begin{center}
                                                \includegraphics[width=2.31in,height=2.22in]{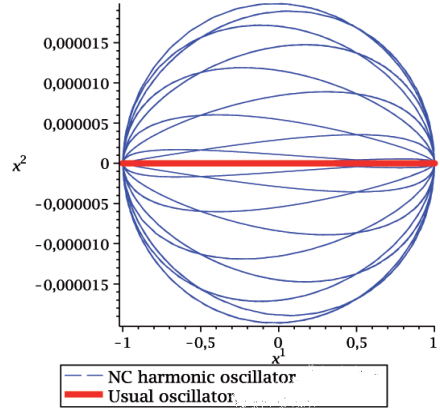}
                                                 %\caption{ Position  $x^{1}$ \textit{versus} $x^{2}$,  for  $t=8\pi$.}
                                                    \end{center}
                                                         \caption{For $t \leq 8\,\pi$}\label{fig3}
                                                      \end{figure}

                              Furthermore, according to the discussion  about the  periodicity conditions  
of  the  NCHO solution carried out in subsection (\ref{APP}), choosing $ \omega = 10 $ and $ 2\Omega =  3$, which are  
{\it relatively prime} numbers, implies that the  NC harmonic   period  is  equal to $ 2\pi $. This result is in accordance  with the trajectories shown in figures \ref{fig2} and \ref{fig3}.
In fact,  in the first figure we note that the NC oscillator  has completed one period  for $ t = 2\pi $, while in the second one,    it has accomplished  four periods for $ t = 8 \pi $. Furthermore, we note that the usual, or commutative, HO does not have velocity towards the $x^{2}$ direction.

                              The distance between the solutions $\vec{r}(t)=\bigg{(}x^{1}(t),x^{2}(t)\bigg{)}$ and  $\vec{r}_{0}(t)=\bigg{(}x^{1}_{0}(t),x^{2}_{0}(t)\bigg{)}$ is defined as 
                               \begin{equation}
                              d\big{(}\vec{r}(t), \vec{r}_{0}(t)\big{)}=||\vec{r}(t)-\vec{r}_{0}(t||=\bigg{(} |x^{1}(t)-x^{1}_{0}(t)|^{2}+|x^{2}(t)-x^{2}_{0}(t)|^{2}\bigg{)}^{\frac{1}{2}}.
                               \end{equation}
                              Using  the same initial conditions  given by  $\vec{r}(0)={(}x^{1}_{00},0{)}$ and $\vec{r}_{0}(0)={(}x^{1}_{00},0{)}$  the distance $d\big{(}\vec{r}(t), \vec{r}_{0}(t)\big{)}$ can be written as

                               \begin{equation}
                                  ||\vec{r}(t)-\vec{r}_{0}(t||=\bigg{(} |\frac{x^{1}_{00}(\alpha\omega)^{2}}{1-(\alpha\omega)^{2}}|^{2} |(1-\cos 2\Omega t)\cos \omega  t|^{2}+
                                   |\frac{x^{1}_{00}(\alpha\omega)}{1-(\alpha\omega)^{2}}|^{2} |(1-\cos 2\Omega t) \sin \omega  t|^{2}\bigg{)}^{\frac{1}{2}},
                                   \end{equation}
                                  which satisfies the following condition
                                  \begin{equation}\label{importante}
 										||\vec{r}(t)-\vec{r}_{0}(t)||\leq \frac{2|\alpha\omega|\sqrt{1+(\alpha\omega)^{2}}}{|1-(\alpha\omega)^{2}|}|x^{1}_{00}|,	
 										\end{equation} 
where we have used the conditions given by
                              \begin{equation}
                              |x^{1}(t)-x^{1}_{0}(t)|=\frac{|x^{1}_{00}(\alpha\omega)^{2}|}{|1-(\alpha\omega)^{2}|} |(1-\cos 2\Omega t)\cos \omega  t|\leq \frac{2|\alpha\omega|^{2}   |x^{1}_{00}|}{|1-(\alpha\omega)^{2}|},
                              \end{equation}
                              \begin{equation}
                               |x^{1}(t)-x^{1}_{0}(t)|=\frac{|x^{1}_{00}(\alpha\omega)|}{|1-(\alpha\omega)^{2}|} |(1-\cos 2\Omega t)\sin \omega  t|\leq \frac{2|\alpha\omega|\,   |x^{1}_{00}|}{|1-(\alpha\omega)^{2}|}.
                                   \end{equation}
                              
The  condition represented in Eq. (\ref{importante}), shows that in some sense the NCY induces a stable perturbation into the usual oscillator.  For all $\epsilon >0 $ there exists a $\delta>0$ such that  $ ||\vec{r}(t)-\vec{r}_{0}(t||\leq \epsilon $ for all $t \in \mathbb{R}$, provide that $|\alpha|< \delta$, where $|\theta^{ij}|=|D^{ij}|$  and $\alpha=\frac{mD^{12}}{2}$. In other words, the r.h.s. of (\ref{importante}) vanishes as $\alpha \longrightarrow\: 0$.  Thus, the difference       between the usual oscillator and its NC counterpart may not be noticeable when the NC parameter $D^{12}$ is sufficiently small. On the other hand, the NC effect becomes more significant and evident as $|x^{1}_{00}|$ increases, since  the distance between the solutions  and the corresponding displacement towards $x^{2}$ direction are   proportional to    $|x^{1}_{00}|$.
  
\begin{figure}[H]
                				          \begin{center}
                                       \includegraphics[width=2.31in,height=2.22in]{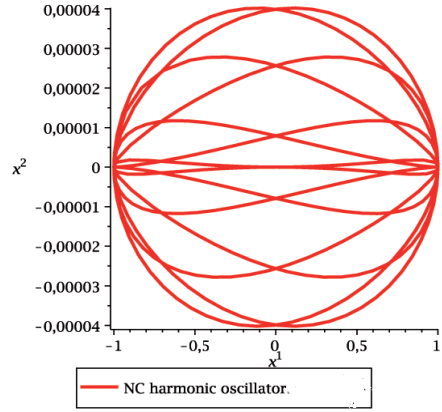}
                                       %\caption{  Position $x^{1}$ \textit{versus} $x^{2}$,  for  $t=\frac{2\pi}{3}$.}
                                       \end{center}
                                       \caption{For $t \leq \frac{2\pi}{3}$}\label{fig4}
                                       \end{figure}

From the solutions shown in Eqs.(\ref{eq:sf333}) and (\ref{eq:sf444}), if the NC oscillator passes through the origin then we must have $\cos( \omega t)=0$ and $\cos ( 2 \Omega t)=1$  for some $t>0$, from these we have that
 \begin{equation}\label{relac}
\frac{2\Omega}{\omega}=\frac{4l}{2k+1},
 \end{equation}  where $k$  and $l$ are positive integers and we  have used that $|\alpha \omega|<1$. Therefore, the solution of the NC oscillator is periodic since $2\Omega / \omega$ is a rational number, i.e., if this NC oscillator passes by the origin then it is necessarily periodic. In order to illustrate this result, the solutions given by  Eqs.(\ref{eq:sf333}) and (\ref{eq:sf444})  are  shown in figure (\ref{fig4}), where we have chosen $\alpha=10^{-6}$,  $\omega=21$, and $\Omega=6$ or,  equivalently, $k=5$ and $l=3$, which is in accordance with (\ref{relac}).
Using  the frequencies chosen  above,  from Eq. (\ref{CP1}), the  period of oscillation  of NC HO  is given by $$T=\frac{2\pi}{\omega}q=\frac{2\pi}{3}.$$ It is  in  agreement with the trajectory shown in figure (\ref{fig4}).

In order to investigate the periodicity conditions of the solutions of $2$-D NC HO,  let us suppose
that the solutions  given by Eqs.(\ref{eq:sf333}) and (\ref{eq:sf444}) are periodic if and only if $\cos( \omega T)=1$  and $\cos( 2\Omega T)=1$, which is equivalent to $2\Omega /\omega$ being a rational number, cf. subsection (\ref{APP}). In fact, let us suppose that the solution given by  Eqs.(\ref{eq:sf333}) and (\ref{eq:sf444}) is periodic, that is, there is a $T>0$ such that 

                             \begin{eqnarray}\label{eq:sf33333}
                                                 0\neq x^{1}(0)= x^{1}(T)&=&\frac{x^{1}_{00}}{1-(\alpha\omega)^{2}}
                                                  \bigg{\{}1- (\alpha\omega)^{2} \cos(2\Omega T)\bigg{\}}  \cos(\omega T),         \\\nonumber
                                                               \end{eqnarray}
               \begin{eqnarray}\label{eq:sf44444}
                                                                0=x^{2}(0)=x^{2}(T)&=&\frac{\alpha\omega x^{1}_{00}}{1-(\alpha\omega)^{2}}
                                                                \bigg{\{}1-  \cos(2\Omega T)\bigg{\}}  \sin(\omega T).       \\\nonumber
                                                                             \end{eqnarray}
From Eq.(\ref{eq:sf44444}) it follows that $\sin( \omega T)=0$ or $\cos( 2\Omega T)=1$. If $\sin( \omega T)=0$ then using  Eq.(\ref{eq:sf33333}) we conclude necessarily that $\cos( \omega T)=1$ for $1-(\alpha\omega)^{2}>0$ and $1-(\alpha\omega)^{2}\cos(2\Omega T)>0$ since $0<|\alpha\omega|<1$. Moreover, according to Eq.(\ref{eq:sf33333}),  $\cos( 2\Omega T)=1$ due to the  assumption that $|\alpha\omega|>0$. If we had the case  $\cos( 2\Omega T)=1$,  Eq.(\ref{eq:sf33333}) also yields the same conclusion $\cos( \omega T)=1$ as claimed before.

Therefore, having in mind the present  hypothesis  $|\alpha\omega|>0$, if $\frac{2\Omega}{\omega}$  is an irrational number
 then the solution given by Eqs.(\ref{eq:sf333}) and (\ref{eq:sf444}) is not periodic.

      Now let us discuss the rotational invariance of the $2$-D NC HO  solutions given by Eqs.(5.14) and (5.15). For this purpose, let us initially consider an arbitrary spatial rotation  of the angle $\beta$ about the origin of the NC HO  initial conditions,  ${x}^{1}_{00},  {x}^{2}_{00}, \dot{x}^{1}_ {00}$, and $\dot{x}^{2}_{00}$;
           
           $$\begin{array}{ccccc}
           \left (
           \begin{array}{c}
                      \tilde{x}^{1}_{00}\\
                      \tilde{x}^{2}_{00}\\
            \end{array}
              \right ) &= &\left (
           \begin{array}{cc}
           \cos(\beta) & -\sin(\beta) \\
           \sin(\beta) & \cos(\beta) \\
            \end{array}
           \right )&.&\left (
                      \begin{array}{c}
                      {x}^{1}_{00}  \\
                       {x}^{2}_{00}\\
                       \end{array}
                      \right ),
            \end{array}$$
             $$\begin{array}{ccccc}
                       \left (
                       \begin{array}{c}
                                  \dot{\tilde{x}}^{1}_{00}\\
                                  \dot{\tilde{x}}^{2}_{00}\\
                        \end{array}
                          \right ) &= &\left (
                       \begin{array}{cc}
                       \cos(\beta) & -\sin(\beta) \\
                       \sin(\beta) & \cos(\beta) \\
                        \end{array}
                       \right )&.&\left (
                                  \begin{array}{c}
                                  \dot{{x}}^{1}_{00}  \\
                                   \dot{{x}}^{2}_{00}\\
                                   \end{array}
                                  \right ).
                        \end{array}$$
%                        \begin{eqnarray}
%                        \tilde{x}^{1}_{00}=\bigg{(}\cos(\beta) {{x}}^{1}_{00}  -\sin(\beta){{x}}^{2}_{00}\bigg{)}\\
%                        \tilde{x}^{2}_{00}=\bigg{(}\sin(\beta){{x}}^{1}_{00}+  \cos(\beta){{x}}^{2}_{00}\bigg{)}\\
%                        \dot{\tilde{x}}^{1}_{00}=\bigg{(}\cos(\beta)\dot{{x}}^{1}_{00}  -\sin(\beta)\dot{{x}}^{2}_{00}\bigg{)}\\
%                        \dot{\tilde{x}}^{2}_{00}=\bigg{(}\sin(\beta)\dot{{x}}^{1}_{00}+ \cos(\beta)\dot{{x}}^{2}_{00}\bigg{)}\\
%                        \end{eqnarray}
                           Here $\tilde{x}^{1}_{00},\tilde{x}^{2}_{00}, \dot{\tilde{x}}^{1}_{00}$, and $ \dot{\tilde{x}}^{2}_{00} $ are the new initial conditions. Substituting these new conditions in Eqs.(\ref{eq:sf33}) and (\ref{eq:sf4}), the NCHO solution $\vec{\tilde{r}}=(\tilde{x}^{1}(t), \tilde{x}^{2}(t))$  may be written as follow
%     \begin{eqnarray}
%     \tilde{x}^{1}=\frac{ x^{1}_{00}}{1-(\alpha\omega)^{2}}\bigg{\{}\cos(\omega t)\cos(\beta)-\alpha\omega \sin(\omega t)\sin(\beta)\\\nonumber
%     +\alpha\omega \sin(\omega t)\sin(\beta)\cos(2\Omega t)
%     -(\alpha\omega)^{2}\cos(\omega t)\cos(2\Omega t)\cos(\beta)\bigg{\}},
%     \end{eqnarray}
%     \begin{eqnarray}
%          \tilde{x}^{2}=\frac{ x^{1}_{00}}{1-(\alpha\omega)^{2}}\bigg{\{}\cos(\omega t)\sin(\beta)+\alpha\omega \sin(\omega t)\cos(\beta)\\\nonumber
%          -\alpha\omega \sin(\omega t)\cos(\beta)\cos(2\Omega t)
%          -(\alpha\omega)^{2}\cos(\omega t)cos(2\Omega t)\sin(\beta)\bigg{\}}.
%          \end{eqnarray}

\begin{eqnarray}\label{eq:sf33a1}
    &&   \tilde{x}^{1}(t)
           =\bigg{\{}\frac{\alpha \sin(\beta)\dot{{x}}^{1}_{00}+ \alpha\cos(\beta)\dot{{x}}^{2}_{00}+ \cos(\beta) {{x}}^{1}_{00}  -\sin(\beta){{x}}^{2}_{00}}{1-(\alpha\omega)^{2}}\bigg{\}}\cos(\omega t) \nonumber \\
           &-&\frac{1}{\omega}\bigg{\{}\frac{-\alpha \omega^{2} \sin(\beta){{x}}^{1}_{00}-\alpha \omega^{2}  \cos(\beta){{x}}^{2}_{00}+ \cos(\beta)\dot{{x}}^{1}_{00}  -\sin(\beta)\dot{{x}}^{2}_{00}}{(\alpha\omega )^{2}-1}\bigg{\}}\sin(\omega t) \nonumber \\
   &+&         \alpha\omega \bigg{\{}\bigg{(}\frac{-\alpha \cos(\beta) \dot{{x}}^{1}_{00}  +\alpha \sin(\beta)\dot{{x}}^{2}_{00}+ \sin(\beta){{x}}^{1}_{00}+  \cos(\beta){{x}}^{2}_{00}}{1-(\alpha\omega )^{2}}\bigg{)} \sin(\omega t) \nonumber \\
  &+&         \frac{1}{\omega}\bigg{(}\frac{\alpha \omega^{2}   \cos(\beta) {{x}}^{1}_{00}  -\alpha \omega^{2}\sin(\beta){{x}}^{2}_{00}+ \sin(\beta)\dot{{x}}^{1}_{00}+ \cos(\beta)\dot{{x}}^{2}_{00}}{(\alpha\omega )^{2}-1}\bigg{)}\cos(\omega t)\bigg{\}}\cos(2\Omega t) \nonumber \\
           &=&\cos(\beta){{x}}^{1}(t)-\sin(\beta)x^{2}(t),\\\nonumber
                    \end{eqnarray}
              \begin{eqnarray}\label{eq:sf4a1}
  &&        \tilde{x}^{2}(t)
               =\bigg{\{}\frac{-\alpha \cos(\beta)\dot{{x}}^{1}_{00}  +\alpha \sin(\beta)\dot{{x}}^{2}_{00}+ \sin(\beta){{x}}^{1}_{00}+  \cos(\beta){{x}}^{2}_{00}}{1-(\alpha\omega )^{2}}\bigg{\}}\cos(\omega t) \nonumber \\
               &-&\frac{1}{\omega}\bigg{\{}\frac{\alpha \omega^{2}   \cos(\beta) {{x}}^{1}_{00}  -\alpha \omega^{2}\sin(\beta){{x}}^{2}_{00}+ \sin(\beta)\dot{{x}}^{1}_{00}+ \cos(\beta)\dot{{x}}^{2}_{00}}{(\alpha\omega )^{2}-1}\bigg{\}}\sin(\omega t) \nonumber \\
               &-&  \alpha\omega \bigg{\{}\bigg{(}\frac{\alpha  \sin(\beta)\dot{{x}}^{1}_{00}+\alpha \cos(\beta)\dot{{x}}^{2}_{00}+ \cos(\beta) {{x}}^{1}_{00}  -\sin(\beta){{x}}^{2}_{00}}{1-(\alpha\omega )^{2}}\bigg{)} \sin(\omega t) \nonumber \\
  &+&             \frac{1}{\omega}\bigg{(}\frac{-\alpha \omega^{2}   \sin(\beta){{x}}^{1}_{00}-\alpha \omega^{2} \cos(\beta){{x}}^{2}_{00}+ \cos(\beta)\dot{{x}}^{1}_{00}  -\sin(\beta)\dot{{x}}^{2}_{00}}{(\alpha\omega )^{2}-1}\bigg{)}\cos(\omega t)\bigg{\}}\cos(2\Omega t) \nonumber \\
                        &=&\sin(\beta){{x}}^{1}(t)+\cos(\beta)x^{2}(t),\\\nonumber
               \end{eqnarray}
                        which  can be rewritten  in a matrix  form as 
     $$\begin{array}{ccccc}\label{fin}
                \left (
                \begin{array}{c}
                           \tilde{x}^{1}(t)\\
                           \tilde{x}^{2}(t)\\
                 \end{array}
                   \right ) &= &\left (
                \begin{array}{cc}
                \cos(\beta) & -\sin(\beta) \\
                \sin(\beta) & \cos(\beta) \\
                 \end{array}
                \right )&.&\left (
                           \begin{array}{c}
                           {x}^{1}(t)  \\
                           x^{2}(t)\\
                            \end{array}
                           \right ).
                 \end{array}$$
                      From this equation, we can conclude   that to carry out a rotation of the HO initial conditions   is equivalent to perform the same  rotation of  the NCHO solution. This result was expected since the NCHO solution given by  Eq.(\ref{eq:sf}) is rotational invariant.

                          From the NC contributions to the  dynamics of the  rotational invariant     HO in an extended phase-space that were observed, we associate the NCY effect with the  oscillatory effect that  varies with  time   and    position. In addition, the intensity and oscillation  frequency  of this field  is  directly related to the modulus of $D^{ij}$  and to the frequency $\Omega $ of  the NC coordinate $\theta^{ij}$.    Another important consideration about this oscillatory effect  regards  the way how it interacts with the  system. Actually, the   effect of this oscillation    explicitly depends on  the linear momentum of the system. In fact,   the  NCY effect on  the motion of a free particle  in NC classical and quantum phase  space  has been associated with the  magnetic field effect in (\cite{18} and references therein), respectively.

\section{Conclusions}
\label{section6}

The analysis of theoretical models described in NC phase-space had brought great attention in this century since Seiberg and Witten \cite{3} discovered that the algebra of N-S string embedded in a magnetic background is NC.  Since string theory is one of the candidates to quantize gravitation it is natural to expect that the geometry of the early Universe (where, theoretically, QM and gravity co-exist) is a NC one.  Since the connection between classical and quantum mechanics is to substitute Poisson brackets by commutators, the investigation of classical mechanics in NC phase-space is interesting.  It is with this relevance that this paper is concerned.

In this work, using firstly a generalized potential and after that a specific potential for a rotational invariant HO we could discuss two formulations for NC theories: the Doplicher-Fredenhagen-Roberts formalism and the minimal canonical DFR extension.  We have obtained the NSL for the NCHO in both ones.

We have constructed an  invariant rotational NSL in a classical phase-space which we assumed to possess  a symplectic structure consistent with the  commutation rules  of  the NCQM in the  extended Hilbert space  that maintains the  rotational symmetry. In this extended Hilbert space, the object of NCY are the coordinate $\theta^{ij} $ and its canonical conjugate momentum $\pi_{ij}$, and both can be considered as operators in a NCQM approach. This formalism considered a canonical extension of DFR approach.  The extended NC Hilbert space is  essential  to recover invariance under rotations and to  construct rotational invariant  NC theories. 

The NC corrections of the new NSL are dependent of:   the variations in the potential, the  NC coordinate $\theta^{ij}$ and its conjugate momentum $\pi_{ij}$. In addition, this  invariant rotational NSL is a generalization of the  NSL  non-invariant under rotations  obtained in other studies  where  $\theta^{ij}$ is  just considered a constant parameter  of NCY \cite{11,13}, i.e., the canonical NC approach. We have applied this NSL in this NC phase-space to treat an  HO  described by   the rotational invariant NC  Hamiltonian in the extended phase-space. In this case, we have obtained the equation of motion for all  phase-space coordinates, including the NC coordinates $\theta^{ij}$, with their respective solutions. The periodicity  conditions  of these solutions were analyzed where we showed that the  solutions of the system are periodic if the ratio  between the  oscillation frequencies is a rational number. In this case, the period is given by $T =\frac {2 \pi} {\omega} m$.

Our investigation looks for more details in the $2$-D NCHO for which we have obtained the solutions that depends only on the initial positions and velocities. The result enables us to compare them, with the aid of graphics, with the solutions of a commutative HO presenting the same initial configuration. We showed that the NC HO solution is periodic if and only if the ratio  $2\Omega /\omega$ is a rational number. 
 The NC oscillator crosses the $x^{2}$-axis and the usual HO passes through the origin simultaneously. Moreover, these are only moments at which the NC oscillator passes through  the $x^{2}$-axis provided that the modulus of the NC coordinate $D^{ij}$ satisfies the relation  $|\alpha \omega|<1$. In addition, the NC oscillator also  passes through the origin if the frequencies satisfy the relation $\frac{2\Omega}{\omega}=\frac{4l}{2k+1}$,  where $k$  and $l$ are positive integers and the fraction is a rational number. Thus, if the NC oscillator passes through the origin, it is necessarily periodic.

 We have obtained  the distance $d\big{(}\vec{r}(t), \vec{r}_{0}(t)\big{)}$ between the $2$-D NC and the commutative oscillator solutions, which  shows that NCY induces a stable perturbation into the ordinary oscillator.  The difference between the usual oscillator and its NC version may be almost unnoticed when the modulus of the NC coordinate is sufficiently small. On the other hand, the  NC effect becomes more significant and noticeable when the initial position $x^{1}_{00}$ increases, since the  distance  between the solutions  and the corresponding displacement towards $x^{2}$ direction are   proportional to    $|x^{1}_{00}|$.

Based on the  NCY effects  observed here,  we have associated the NCY to an oscillatory effect which varies with time and space. In this case,  the intensity of this oscillation is directly related to the modulus of  the NC coordinate $\theta^{ij}$ and its   effective interaction   with the system depends explicitly on the  linear momentum of the system. 
  
From all that was accomplished here,  it is worth to emphasize the fact that we have explored the dynamics of  the NC  parameter $ \theta^{ij} $. Here,   considered as a coordinate of phase-space,  its  dynamics was completely described for the case of the NCHO. However,  we understand that the dynamics of the  NC coordinates  $ \theta^{ij} $ deserves further studies in other physical systems of interest.

Finally, through the analysis of the equations of motion for a generalized potential and for the NCHO we have obtained that the momentum associated to the coordinate $\theta^{ij}$ is in fact essential to construct the DFR phase-space.  This fact happens because we saw that when $\pi_{ij}=0$ we have $\theta^{ij}=const.$.  Hence, we have recovered the canonical NCY.  The phase-space reduction obeys the representation $(x^i,p_i, \theta^{ij},\pi_{ij}) \longrightarrow (x^i,p_i)$ and not $(x^i, p_i, \theta^{ij})$ which would be natural to expect.  We believe that this result makes the momentum $\pi_{ij}$ relevant, contrarily to the current literature.  Concerning QFT, this result is more important because $\theta^{\mu\nu} \not= const.$ is connected to Lorentz invariance.  Since we have $\pi_{\mu\nu}=0$ the Lorentz invariance is broken, which makes, again, the momentum $\pi_{\mu\nu}$ a relevant quantity.  We conclude that the Lorentz invariance is connected not only to the variable feature of $\theta^{\mu\nu}$ but also to $\pi_{\mu\nu}$.  This momentum relevance in DFR formalism  is new in the literature.

\acknowledgments  The authors would like to thank CNPq (Conselho Nacional de Desenvolvimento Cient\' ifico e Tecnol\'ogico) and FAPEMIG (Funda\c{c}\~ao de Amparo \`a Pesquisa do Estado de Minas Gerais), brazilian scientific support agencies, for partial financial support.

%\lllll

\end{document}